\documentclass[prd,twocolumn,showpacs,linenumbers,groupedaddress,superscriptaddress,amsmath,amssymb]{revtex4}

\usepackage{graphicx}
\usepackage{dcolumn}
\usepackage{bm}
\usepackage{amssymb}
\usepackage{enumerate}
\usepackage{lineno}
\usepackage{multirow} 

\newcommand\ainv{A'\to invisible}

\newcommand\g{\gamma}
\newcommand\ma{m_{A'}}

\newcommand\Na{{N}_{A'}}

\newcommand\emu{e^- Z \to e^- Z \g; \g \to \mu^+ \mu^-}

\def\address{\@ifstar{\address@star}%
  {\@ifnextchar[{\address@optarg}{\address@noptarg}}}

\begin{document}

\title{ Dark Matter Search in Missing Energy Events with NA64
}
\affiliation{\it Universit\"at Bonn, Helmholtz-Institut f\"ur Strahlen-und Kernphysik, 53115 Bonn, Germany} 
\affiliation{\it  Joint Institute for Nuclear Research, 141980 Dubna, Russia}
\affiliation{\it Technische Universit\"at M\"unchen, Physik  Department, 85748 Garching, Germany}
\affiliation{\it CERN, European Organization for Nuclear Research, CH-1211 Geneva, Switzerland}
\affiliation{\it University of Illinois at Urbana Champaign, Urbana, 61801-3080 Illinois, USA}
\affiliation{\it Department of Physics and Astronomy, University College London, Gower St., London WC1E 6BT, United Kingdom}
\affiliation{\it Institute for Nuclear Research, 117312 Moscow, Russia} 
\affiliation{\it P.N. Lebedev Physical Institute, Moscow, Russia, 119 991 Moscow, Russia}
\affiliation{\it Skobeltsyn Institute of Nuclear Physics, Lomonosov Moscow State University, 119991  Moscow, Russia}
\affiliation{\it Physics Department, University of Patras, 265 04 Patras, Greece} 
\affiliation{\it State Scientific Center of the Russian Federation Institute for High Energy Physics of National Research Center 'Kurchatov Institute' (IHEP), 142281 Protvino, Russia}
\affiliation{\it  Departamento de Ciencias F\i{i}sicas, Universidad Andres Bello, Sazi\'{e} 2212, Piso 7, Santiago, Chile}
\affiliation{\it Tomsk State Pedagogical University, 634061 Tomsk, Russia}
\affiliation{\it Universidad T\'{e}cnica Federico Santa Mar\'{i}a, 2390123 Valpara\'{i}so, Chile}
\affiliation{\it ETH Z\"urich, Institute for Particle Physics and Astrophysics, CH-8093 Z\"urich, Switzerland}
\author{D.~Banerjee}\affiliation{\it CERN, European Organization for Nuclear Research, CH-1211 Geneva, Switzerland}\affiliation{\it University of Illinois at Urbana Champaign, Urbana, 61801-3080 Illinois, USA}
\author{V.~E.~Burtsev}\affiliation{\it  Joint Institute for Nuclear Research, 141980 Dubna, Russia}
\author{A.~G.~Chumakov}\affiliation{\it Tomsk State Pedagogical University, 634061 Tomsk, Russia}
\author{D.~Cooke}\affiliation{\it Department of Physics and Astronomy, University College London, Gower St., London WC1E 6BT, United Kingdom}
\author{P.~Crivelli}\affiliation{\it ETH Z\"urich, Institute for Particle Physics and Astrophysics, CH-8093 Z\"urich, Switzerland}
\author{E.~Depero}\affiliation{\it ETH Z\"urich, Institute for Particle Physics and Astrophysics, CH-8093 Z\"urich, Switzerland}
\author{A.~V.~Dermenev}\affiliation{\it Institute for Nuclear Research, 117312 Moscow, Russia}
\author{S.~V.~Donskov}\affiliation{\it State Scientific Center of the Russian Federation Institute for High Energy Physics of National Research Center 'Kurchatov Institute' (IHEP), 142281 Protvino, Russia}
\author{R.~R.~Dusaev}\affiliation{\it Tomsk State Pedagogical University, 634061 Tomsk, Russia}
\author{T.~Enik}\affiliation{\it  Joint Institute for Nuclear Research, 141980 Dubna, Russia}
\author{N.~Charitonidis}\affiliation{\it CERN, European Organization for Nuclear Research, CH-1211 Geneva, Switzerland}
\author{A.~Feshchenko}\affiliation{\it  Joint Institute for Nuclear Research, 141980 Dubna, Russia}
\author{V.~N.~Frolov}\affiliation{\it  Joint Institute for Nuclear Research, 141980 Dubna, Russia}
\author{A.~Gardikiotis}\affiliation{\it Physics Department, University of Patras, 265 04 Patras, Greece}
\author{S.~G.~Gerassimov }\affiliation{\it P.N. Lebedev Physical Institute, Moscow, Russia, 119 991 Moscow, Russia}\affiliation{\it Technische Universit\"at M\"unchen, Physik  Department, 85748 Garching, Germany}
\author{S.~N.~Gninenko}\affiliation{\it Institute for Nuclear Research, 117312 Moscow, Russia}
\author{M.~H\"osgen}\affiliation{\it Universit\"at Bonn, Helmholtz-Institut f\"ur Strahlen-und Kernphysik, 53115 Bonn, Germany}
\author{M.~Jeckel}\affiliation{\it CERN, European Organization for Nuclear Research, CH-1211 Geneva, Switzerland}
\author{A.~E.~Karneyeu}\affiliation{\it Institute for Nuclear Research, 117312 Moscow, Russia}
\author{G.~Kekelidze}\affiliation{\it  Joint Institute for Nuclear Research, 141980 Dubna, Russia}
\author{B.~Ketzer}\affiliation{\it Universit\"at Bonn, Helmholtz-Institut f\"ur Strahlen-und Kernphysik, 53115 Bonn, Germany}
\author{D.~V.~Kirpichnikov}\affiliation{\it Institute for Nuclear Research, 117312 Moscow, Russia}
\author{M.~M.~Kirsanov}\affiliation{\it Institute for Nuclear Research, 117312 Moscow, Russia}
\author{I.~V.~Konorov}\affiliation{\it P.N. Lebedev Physical Institute, Moscow, Russia, 119 991 Moscow, Russia} \affiliation{\it Technische Universit\"at M\"unchen, Physik  Department, 85748 Garching, Germany}
\author{S.~G.~Kovalenko}\affiliation{\it Universidad T\'{e}cnica Federico Santa Mar\'{i}a, 2390123 Valpara\'{i}so, Chile}
\author{V.~A.~Kramarenko}\affiliation{\it  Joint Institute for Nuclear Research, 141980 Dubna, Russia}\affiliation{\it Skobeltsyn Institute of Nuclear Physics, Lomonosov Moscow State University, 119991  Moscow, Russia}
\author{L.~V.~Kravchuk}\affiliation{\it Institute for Nuclear Research, 117312 Moscow, Russia}
\author{ N.~V.~Krasnikov}\affiliation{\it Institute for Nuclear Research, 117312 Moscow, Russia}
\author{S.~V.~Kuleshov}\affiliation{\it Departamento de Ciencias F\i{i}sicas, Universidad Andres Bello, Sazi\'{e} 2212, Piso 7, Santiago, Chile}
\author{V.~E.~Lyubovitskij}\affiliation{\it Tomsk State Pedagogical University, 634061 Tomsk, Russia}\affiliation{\it Universidad T\'{e}cnica Federico Santa Mar\'{i}a, 2390123 Valpara\'{i}so, Chile}
\author{V.~Lysan}\affiliation{\it  Joint Institute for Nuclear Research, 141980 Dubna, Russia}
\author{V.~A.~Matveev}\affiliation{\it  Joint Institute for Nuclear Research, 141980 Dubna, Russia}
\author{Yu.~V.~Mikhailov}\affiliation{\it State Scientific Center of the Russian Federation Institute for High Energy Physics of National Research Center 'Kurchatov Institute' (IHEP), 142281 Protvino, Russia}
\author{L.~Molina Bueno}\affiliation{\it ETH Z\"urich, Institute for Particle Physics and Astrophysics, CH-8093 Z\"urich, Switzerland}
\author{D.~V.~Peshekhonov}\affiliation{\it  Joint Institute for Nuclear Research, 141980 Dubna, Russia}
\author{V.~A.~Polyakov}\affiliation{\it State Scientific Center of the Russian Federation Institute for High Energy Physics of National Research Center 'Kurchatov Institute' (IHEP), 142281 Protvino, Russia}
\author{B.~Radics}\affiliation{\it ETH Z\"urich, Institute for Particle Physics and Astrophysics, CH-8093 Z\"urich, Switzerland}
\author{R.~Rojas}\affiliation{\it Universidad T\'{e}cnica Federico Santa Mar\'{i}a, 2390123 Valpara\'{i}so, Chile}
\author{A.~Rubbia}\affiliation{\it ETH Z\"urich, Institute for Particle Physics and Astrophysics, CH-8093 Z\"urich, Switzerland}
\author{V.~D.~Samoylenko}\affiliation{\it State Scientific Center of the Russian Federation Institute for High Energy Physics of National Research Center 'Kurchatov Institute' (IHEP), 142281 Protvino, Russia}
\author{D.~Shchukin}\affiliation{\it P.N. Lebedev Physical Institute, Moscow, Russia, 119 991 Moscow, Russia}
\author{V.~O.~Tikhomirov}\affiliation{\it P.N. Lebedev Physical Institute, Moscow, Russia, 119 991 Moscow, Russia}
\author{I.~Tlisova}\affiliation{\it Institute for Nuclear Research, 117312 Moscow, Russia} 
\author{D.~A.~Tlisov}\affiliation{\it Institute for Nuclear Research, 117312 Moscow, Russia} 
\author{A.~N.~Toropin}\affiliation{\it Institute for Nuclear Research, 117312 Moscow, Russia}
\author{A.~Yu.~Trifonov}\affiliation{\it Tomsk State Pedagogical University, 634061 Tomsk, Russia}
\author{B.~I.~Vasilishin}\affiliation{\it Tomsk State Pedagogical University, 634061 Tomsk, Russia}
\author{G.~Vasquez Arenas}\affiliation{\it Universidad T\'{e}cnica Federico Santa Mar\'{i}a, 2390123 Valpara\'{i}so, Chile}
\author{P.~V.~Volkov}\affiliation{\it  Joint Institute for Nuclear Research, 141980 Dubna, Russia}\affiliation{\it Skobeltsyn Institute of Nuclear Physics, Lomonosov Moscow State University, 119991  Moscow, Russia}
\author{V.~Yu.~Volkov}\affiliation{\it Skobeltsyn Institute of Nuclear Physics, Lomonosov Moscow State University, 119991  Moscow, Russia}
\author{P.~Ulloa}\affiliation{\it Universidad T\'{e}cnica Federico Santa Mar\'{i}a, 2390123 Valpara\'{i}so, Chile}

%
%
\collaboration{The NA64 Collaboration}\noaffiliation
\vskip 0.25cm

\date{\today}


\begin{abstract}
 A  search for sub-GeV dark matter production mediated by a new  vector  boson $A'$, called a  dark photon,  is performed by the  NA64
  experiment  in missing energy events from 100 GeV electron interactions in an active beam dump  at the CERN SPS.     
  From the analysis of the data collected in the years  2016, 2017, and 2018 with  $2.84\times10^{11}$  electrons on target  
  no evidence of such a process  has been found. 
The  most  stringent constraints on the $A'$ mixing strength with photons and  the parameter space for the scalar and fermionic 
dark matter in the  mass range  $\lesssim 0.2$ GeV are derived, thus demonstrating the power of the active beam dump approach
for the dark matter search. 
\end{abstract}


\maketitle
 The idea  that in addition to gravity a  new force between the  dark and visible matter transmitted by a vector  boson,  $A'$ , called dark photon, might  exist is quite exciting ~\cite{Fayet,prv,ArkaniHamed:2008qn,jr}. 
The $A'$ can  have a mass in the sub-GeV mass range,  and couple to 
the standard model (SM)  via kinetic mixing  with the ordinary photon,  
described by the  term  $\frac{\epsilon}{2}F'_{\mu\nu}F^{\mu\nu}$ and  parametrized by the mixing strength  $\epsilon$. An example of the Lagrangian of the SM extended by  the dark sector (DS)  is given by:
\begin{eqnarray}
\mathcal{L} = \mathcal{L}_{SM}  -\frac14 F'_{\mu\nu}F'^{\mu\nu}+
\frac{\epsilon}{2}F'_{\mu\nu}F^{\mu\nu}+  \frac{m_{A'}^2}{2}A'_\mu A'^\mu \nonumber \\
+ i \bar{\chi}\gamma^\mu \partial_\mu \chi -m_\chi \bar{\chi} \chi -e_D \bar{\chi}\gamma^\mu A'_\mu \chi, 
\label{DarkSectorLagrangian}
\end{eqnarray}
where the  massive $A'_\mu$  field is associated with the  spontaneously broken 
$U_D(1)$ gauge group, 
$F'_{\mu\nu} = \partial_\mu A'_\nu-\partial_\nu A'_\mu$,  and $m_{A'},~m_\chi$  are, respectively,  the masses of the $A'$  and  dark matter (DM)  particles, $\chi$,   which are  treated as Dirac fermions   coupled to 
$A'_{\mu}$  with  the dark  coupling strength  $e_D$ of the $U(1)_D$ gauge interactions. The mixing term  of ~\eqref{DarkSectorLagrangian}
 results in the interaction $\mathcal{L}_{int}= \epsilon e A'_{\mu} J^\mu_{em}$ 
of dark photons with the electromagnetic current $J^\mu_{em}$ with a strength $\epsilon e$, 
where $e$ is the electromagnetic coupling and $\epsilon \ll 1$ \cite{Okun:1982xi,Galison:1983pa,Holdom:1985ag}. Such small values of $\epsilon$ 
can be obtained in grand unified theories  from loop effects  of particles charged under both the dark $U_D(1)$ and SM  $U(1)$ interactions with  a typical one-loop  value  $\epsilon = e e_D/16\pi^2 \simeq 10^{-2}-10^{-4}$ \cite{Holdom:1985ag}, or from  two-loop contributions resulting  in 
$\epsilon \simeq 10^{-3}-10^{-5}$. The accessibility of these values at accelerator experiments  has motivated a worldwide effort towards  dark forces and other portals between the visible and dark sectors; see Refs. \cite{jr,gk,Fayet:2007ua,Pospelov:2008zw, Essig:2013lka,report1,report2, pbc-bsm, pbc, berlin,  pdg} for a review.
 \begin{figure*}[tbh!!]
\includegraphics[width=.7\textwidth]{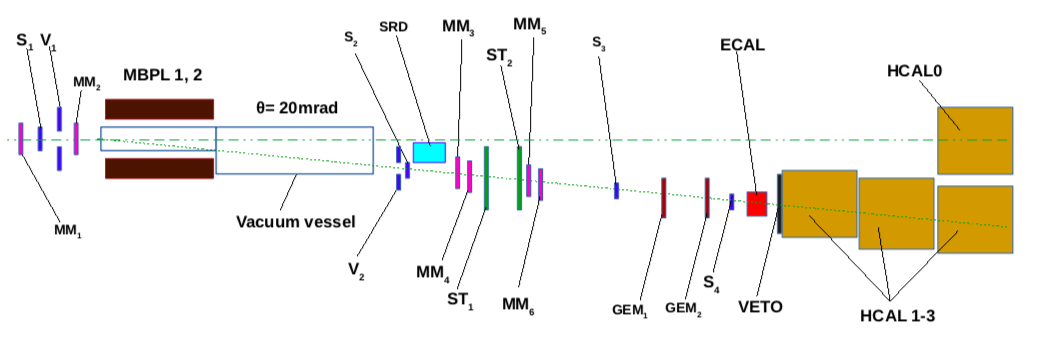}%
\vskip-0.cm{\caption{Schematic illustration of the setup to search for $\ainv$  decays of the bremsstrahlung $A'$s 
produced in the reaction  $eZ\rightarrow eZ A'$ of 100 GeV $e^-$ incident  on the active ECAL target.\label{setup}}}
\end{figure*} 
\par 
If the $A'$ is the lightest state in the dark sector,  then it would decay mainly visibly  to SM leptons $l$ (or hadrons); see, e.g., \cite{apex, merkel, babar1, phenix, na48, kloe3}, and  also \cite{pdg}. 
In the presence of light DM states $\chi$ with the masses $m_\chi<\ma/2$, the $A'$  would predominantly  decay invisibly  into those particles provided that  $e_D >\epsilon e$. 
Various dark sector models motivate the existence of sub-GeV scalar and Majorana or  pseudo-Dirac  DM coupled to the $A'$ 
\cite{report2, pbc-bsm, deNiverville:2011it,pdg,Izaguirre:2014bca,Iza2015,Iza2017,luc}.  To interpret the observed  abundance of DM relic density, 
the requirement of the  thermal freeze-out of DM annihilation into visible matter 
through $\g-A'$  mixing allows one to derive a relation  
\begin{equation}
\alpha_D \simeq 0.02 f \Bigl( \frac{10^{-3}}{\epsilon}\Bigr)^2\Bigl( \frac{\ma}{100~ {\text MeV}}\Bigr)^4
\Bigl( \frac{10~{\text MeV}}{m_\chi}\Bigr)^2
\label{alphad}
\end{equation}
where $\alpha_D = e_D^2/4\pi$   and  the parameter $f$ depends on $m_{A^`}$ and $m_{\chi}$ \cite{Kolb}.
For $\frac{m_{A^`}}{m_{\chi}} =3$,  $f \lesssim 10$ for a scalar \cite{deNiverville:2011it}, and $f\lesssim 1$ for a  fermion \cite{Izaguirre:2014bca}. 
 This  prediction
provides  an important target for the ($\epsilon, ~\ma$) parameter space which can be probed at the CERN SPS energies. 
Models introducing the invisible  $A'$ also  may explain various astrophysical  anomalies \cite{Lee:2014tba} and are subject to various experimental   constraints    leaving,  however,  a  large area that is still unexplored \cite{deNiverville:2011it,Diamond:2013oda,hd,Essig:2013vha, Batell:2009di, e137th, na64prl, minib2018, na64prd,babarg-2,na62}.  
\par In this work  we report new results on the search for  the $A'$ mediator and light dark matter (LDM)  in the fixed-target  experiment NA64 at the CERN SPS.
 In the following we assume  that the $A'$ invisible decay mode is predominant, i.e. $\Gamma (A'\rightarrow \bar{\chi}\chi)/\Gamma_{tot} \simeq 1$.
 If such  invisible $A'$ exists, many crucial questions about its coupling constants, mass scale,  decay modes, etc. arise. One possible way to answer these questions is  to search  for the  $A'$ in fixed-target experiments. The $A'$s  could be 
produced by a high-intensity beam in a dump and generate a flux of DM particles  through the $A'\rightarrow \bar{\chi}\chi$ decay,  which can be detected through the  scattering off electrons in  the far target \cite{deNiverville:2011it,Izaguirre:2014bca,Diamond:2013oda,Batell:2009di, Gninenko:2012prd,Gninenko:2012plb}.   The 
signal event rate  in the detector in this case, scales as $\epsilon^2 y \propto \epsilon^4 \alpha_D$, with one $\epsilon^2$ associated with  the $A'$ production in the dump and 
 $\epsilon^2 \alpha_D$ coming from the $\chi$ particle scattering in the detector, and with  the parameter $y$   defined as 
 $y = \epsilon^2 \alpha_D \Bigl(\frac{m_\chi}{m_{A'}}\Bigr)^4$.
Another method, discussed in this work and proposed in Refs.~\cite{Gninenko:2013rka,Andreas:2013lya}, 
 is based on the detection of the missing energy, carried away by the hard bremsstrahlung $A'$
 produced in the process $e^-Z \to e^- Z A';  \ainv$  of high-energy electrons scattering in the active beam dump target.   The advantage of this type of experiment compared to the beam dump ones  is that its sensitivity is proportional to $\epsilon^2$,  associated with the $A'$ production  and its subsequent prompt invisible decay.
\begin{figure*}[tbh!!]
\includegraphics[width=0.45\textwidth]{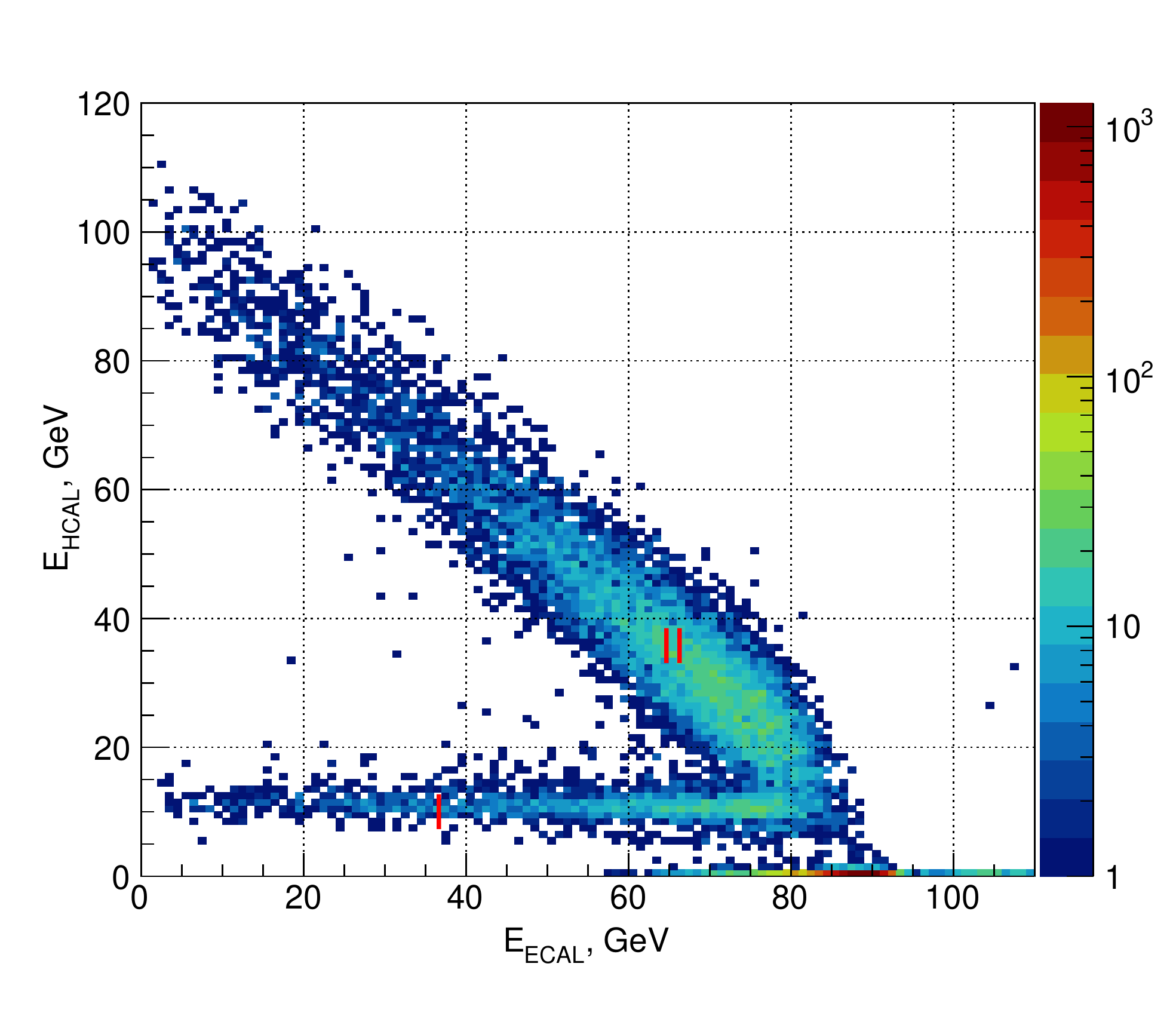}
\hspace{-0.cm}{\includegraphics[width=0.45\textwidth]{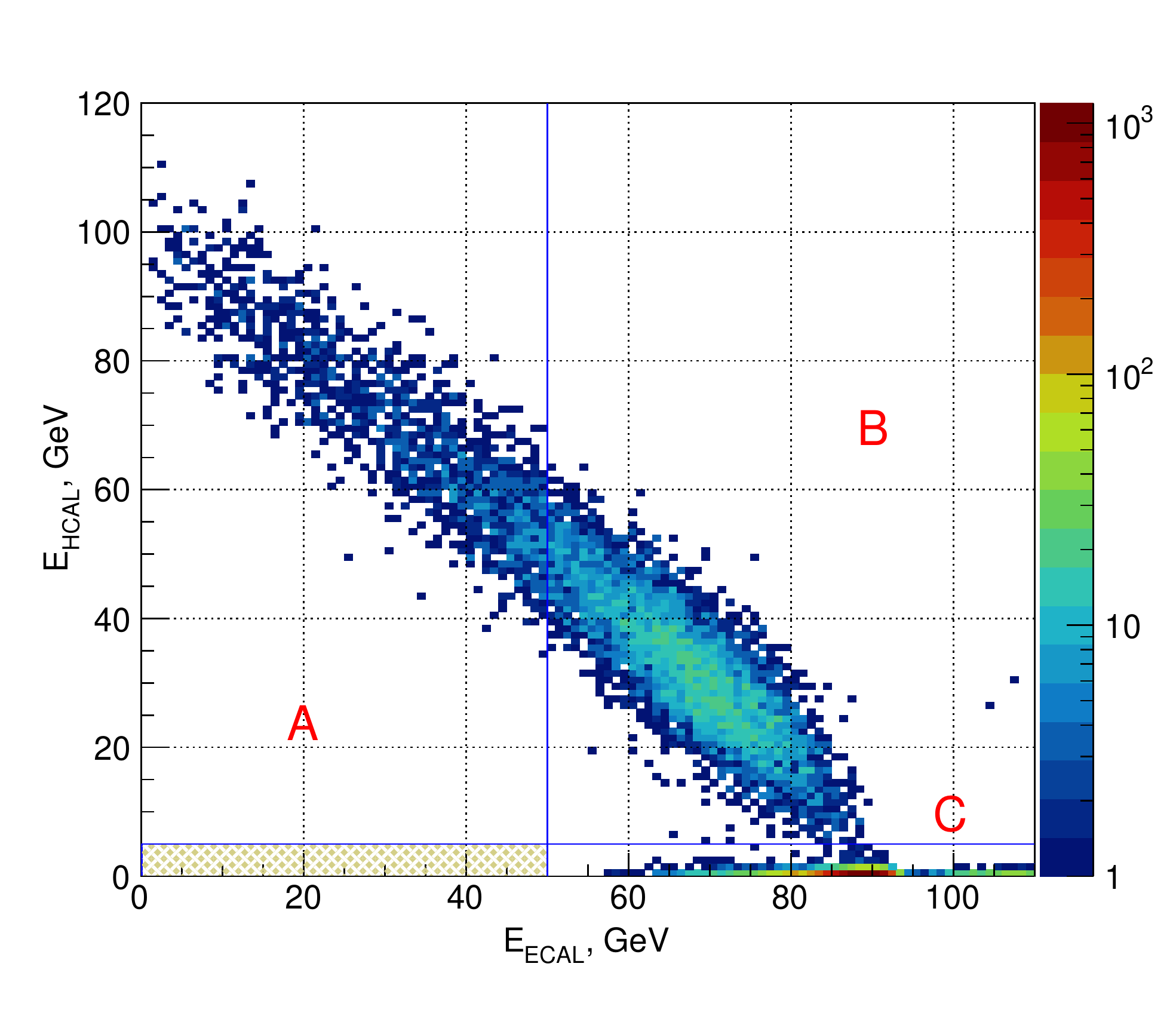}}
\caption{The left panel shows the measured  distribution of events in the ($E_{ECAL}$;$E_{HCAL}$) plane from the combined  run data at the earlier phase of the analysis. The right  panel shows the same distribution  after applying all selection criteria. The shaded area is the signal box,  which contains no events. The size of the signal box along the $E_{HCAL}$ axis is increased by a factor of 5 for illustration purposes. The side bands $A$ and $C$ are the ones  used for the background estimate inside the signal region.  }
\label{ecvshc}
\end{figure*}   
\par  The  NA64 detector  is schematically shown in Fig.~\ref{setup}.
The experiment  employed  the optimized H4  100 GeV  electron beam \cite{h4}.  
The  beam  has a maximal intensity $\simeq 10^7$ electrons per  SPS spill of 4.8 s produced by the primary 400 GeV proton beam  with an intensity of few 10$^{12}$ protons on target.  The detector utilized 
the beam defining  scintillator (Sc)  counters $S_{1-4}$ and veto  $V_{1,2}$, a magnetic spectrometer  consisting of two successive  dipole magnets MBPL$_{1,2}$ with the integral  magnetic field of $\simeq$7 T$\cdot$m  and a low-material-budget tracker. The tracker was a set of two upstream Micromegas chambers MM$_{1,2}$,  and four MM$_{3-6}$,  downstream  stations, as well as two straw-tube  ST$_{1,2}$ and GEM$_{1,2}$ chambers  allowing the measurements of $e^-$ momenta with the precision $\delta p/p \simeq 1\%$ \cite{Banerjee:2015eno}. 
To enhance  electron identification,  synchrotron radiation (SR) emitted in the MBPL magnetic field  
 was used for their efficient  tagging with a SR detector (SRD), which was an array  of a  PbSc sandwich calorimeter of a very fine segmentation \cite{Gninenko:2013rka, na64srd}. By using the SRD the initial admixture  of the hadron contamination in the  beam $\pi/e^- \lesssim 10^{-2}$ was further suppressed by a factor $\simeq 10^3$.   
The detector was also equipped with an active dump target, which is an  electromagnetic calorimeter (ECAL),  a  matrix of $6\times 6 $  Shashlik-type modules  
 assembled from  Pb and Sc plates  for  measurement of the electron energy  $E_{ECAL} $. 
 Each module has $\simeq 40$ radiation  lengths ($X_0$) with the first 4$X_0$ serving as a preshower detector.   
 Downstream of the ECAL, the detector was equipped with a large  high-efficiency veto counter VETO, and a massive, hermetic hadronic calorimeter (HCAL) of $\simeq 30$ nuclear interaction lengths in total. The modules HCAL$_{1-3}$  provided  an efficient veto to detect muons or hadronic secondaries produced in the $e^- A$ interactions  in the target.  
 The events were collected with the hardware trigger  requiring  an in-time cluster in the ECAL with the energy $E_{ECAL} \lesssim 80$ GeV. 
 The search described in this paper uses the data samples of $n_{EOT}=0.43\times 10^{11}, 0.56\times 10^{11}$ and 
 $1.85\times 10^{11}$ electrons on target (EOT),  collected  in  the years 2016, 2017 and 2018  with the beam intensities 
 in the range $\simeq (1.4-6)\times 10^6,~\simeq (5-6)\times 10^6$ and  $\simeq(5-9)\times 10^6$   e$^-$ per spill, respectively.
 Data corresponding it total to $2.84\times 10^{11}$ EOT from these three runs (hereafter called respectively  runs I,II, and III) were processed with  selection criteria similar to the one used in Ref. \cite{na64prd} and finally combined as described below. Compared to the analysis of Ref.\cite{na64prd}, a number of improvements , in particular in the track  reconstruction 
  were made  in the 2018 run to increase the overall efficiency. Also,  the zero-degree calorimeter HCAL$_0$ was used to reject events accompanied by  hard neutrals from the upstream $e^-$ interactions, see Fig.~\ref{setup}.
\par In order to avoid biases in the determination of selection criteria for signal events, a blind analysis was performed. Candidate events were requested  to have the missing energy  $E_{miss} = E_0 - E_{ECAL}> 50 $ GeV. The signal box ($E_{ECAL} < 50~{\text GeV }; E_{HCAL} < 1~{\text GeV }$) was defined based on the energy spectrum calculations for $A'$s emitted by $e^\pm$ from the electromagnetic ($e-m$) shower generated by the primary  $e^-$s in the  target \cite{gkkk, gkkketl}. 
 A  Geant4 \cite{Agostinelli:2002hh, geant} based Monte Carlo (MC) simulation used to study the detector performance,  signal acceptance, and  background level, as well as the analysis procedure  including  selection of  cuts and estimate of the sensitivity  are   described in detail  in Ref.\cite{na64prd}. 
\par The left panel in Fig.~\ref{ecvshc} shows the distribution of $\simeq 3\times 10^4$ events from the reaction 
$e^- Z \to anything$ in the  $(E_{ECAL}; E_{HCAL})$ plane measured with loose selection criteria requiring  mainly the presence of a beam $e^-$ 
identified with the SR tag. Events from area I originate from the QED dimuon production,  dominated by the 
reaction  $\emu$  with   a hard bremsstrahlung photon conversion on a target nucleus and   characterized by  the energy of $ \simeq 10$ GeV deposited by the dimuon pair in the HCAL. This  rare process was used as a benchmark  allowing us  to verify the reliability of the MC simulation, correct  the signal acceptance,  cross-check  systematic uncertainties  and background estimate \cite{na64prd}. Region II shows  the SM events from the hadron electroproduction in the target that satisfy 
the energy conservation $E_{ECAL} + E_{HCAL} \simeq 100$ GeV  within the energy resolution of the detectors. 
\par  Finally, the following   selection criteria  were chosen to maximize the acceptance for  signal events  and to minimize  background.
  (i) The incoming particle track  should have the momentum $100\pm 3$ GeV and a  small angle with respect to  the beam axis  to reject large angle tracks from the upstream $e^-$ interactions. 
(ii)  The energy deposited in the SRD detector should be within the SR range emitted by $e^-$s and in time with the trigger. 
(iii)  The lateral and longitudinal shape of the shower in the ECAL should be  consistent with the  one expected for the signal shower \cite{gkkk}.
(iv) There should be no multiple hits activity in the straw-tube chambers, which was  an effective  cut against  hadron electroproduction in the beam material upstream of the dump, and  no activity in VETO. Only  $\simeq 1.6 \times 10^4 $ events passed these criteria from combined runs.
\begin{table}[tbh!] 
\begin{center}
\caption{Expected background   for   $2.84\times 10^{11}$ EOT.}\label{tab:bckg}
\vspace{0.15cm}
\begin{tabular}{lr}
\hline
\hline
Background source& Background, $n_b$\\
\hline
(i) dimuons &$ 0.024\pm 0.007$\\
(ii)  $\pi,~K\to e \nu $, $K_{e3}$ decays& $ 0.02\pm 0.01$ \\
(iii) $e^-$ hadron interactions in the beam line &$0.43\pm 0.16$\\
(iv) $e^-$ hadron interactions in the target & $<0.044$ \\
(v) Punch-through $\g$'s, cracks, holes & $<0.01$\\
\hline 
Total $n_b$ (conservatively)   &    $0.53\pm 0.17$\\
\hline
\hline 
\end{tabular}
\end{center}
\end{table}
\par There are several  background sources shown in Table \ref{tab:bckg} that may fake the  signal: (i) loss of dimuons due 
to statistical fluctuations of the signal or muon decays,  (ii) decays in flight of mistakenly SRD tagged  
$\pi$, $K$  (iii) the energy loss from the $e^-$ hadronic interactions in the beam line due to the insufficient downstream detector coverage,
and (iv) punch-through of leading neutral hadrons $(n, K^0_L)$ produced in the $e^-$  interactions in the target. 
\begin{figure}[tbh]
\begin{center}
\includegraphics[width=0.45\textwidth]{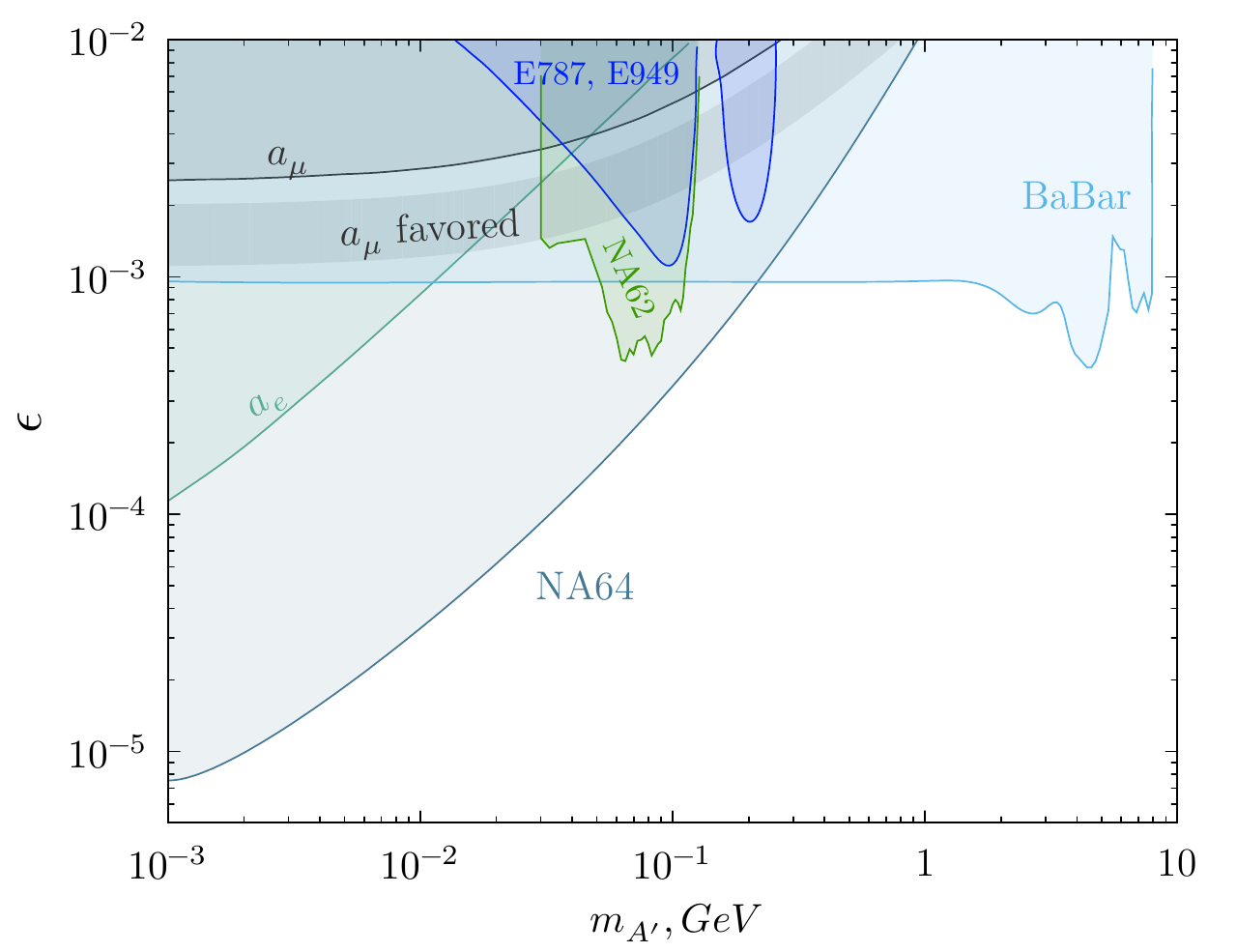}
\caption {The NA64 90\% C.L. exclusion region in the ($m_{A'}, \epsilon$) 
plane.   Constraints from the  
 E787 and  E949  \cite{hd,Essig:2013vha}, $BABAR$ \cite{babarg-2} and recent NA62 \cite{na62} experiments,  as well as the muon  $\alpha_\mu$ favored area 
 are also shown. 
 For more limits from indirect searches and 
 planned measurements; see,  e.g.,  Refs. \cite{report1,report2, pbc-bsm}.
  \label{exclinv}}
\end{center}
\end{figure} 
   The   backgrounds (i) and (ii)  were  simulated with the full statistics of the data. 
 The background estimate in the case (iii) was mainly obtained  from data by the extrapolation of events  from the sideband 
 $C$ ($E_{ECAL} > 50~{\text GeV }; E_{HCAL} < 1~{\text GeV }$) shown in the right panel of Fig. \ref{ecvshc} into the signal region and  assessing the systematic errors by varying the fit functions selected as described in Ref. \cite{na64prd}. The shape of the extrapolation functions  was  taken from the analysis of a much larger data sample of events from case (iv),  and 
 cross-checked with  simulations of the $e^-$  hadronic interactions in the dump.  For case (iv),  events from the region A ($E_{ECAL} < 50~{\text GeV }; E_{HCAL} > 1~{\text GeV }$) of Fig. \ref{ecvshc},  which  are pure neutral hadronic secondaries produced in the ECAL, were used.
The background (iv)  was extracted from the data themselves  by using  the longitudinal segmentation of HCAL for the  conservative punch-through probability estimate. 
   After determining all the selection criteria and background levels, we unblind the data.  No event in the signal box was found, as shown  in  Fig.~\ref{ecvshc}, allowing  us to obtain the $m_{A'}$-dependent upper limits on the mixing strength. 
\par In the final combined statistical analysis,  runs I-III were analysed simultaneously using the multibin limit setting technique \cite{na64prd}
 based on the RooStats package \cite{root}. First, the background estimate, efficiencies,  and their  corrections and uncertainties were used to optimize the main cut defining the signal box,  by comparing
sensitivities,  defined as an average expected limit calculated using the profile likelihood method. The calculations were done  
with uncertainties used as nuisance parameters, assuming their  log-normal distributions \cite{Gross:2007zz}. For this optimization, 
the most important inputs were the expected values from the background extrapolation into the signal region from  the data 
samples of  runs I-III with their errors estimated from the  variation of 
the extrapolation functions. The optimal  cut was found to be  weakly dependent on the $A'$ mass choice and can be safely set to $E_{ECAL} \lesssim 50$ GeV
for the whole mass range.  
\begin{figure*}[tbh!]
\hspace{-0.5cm}{\includegraphics[width=.44\textwidth]{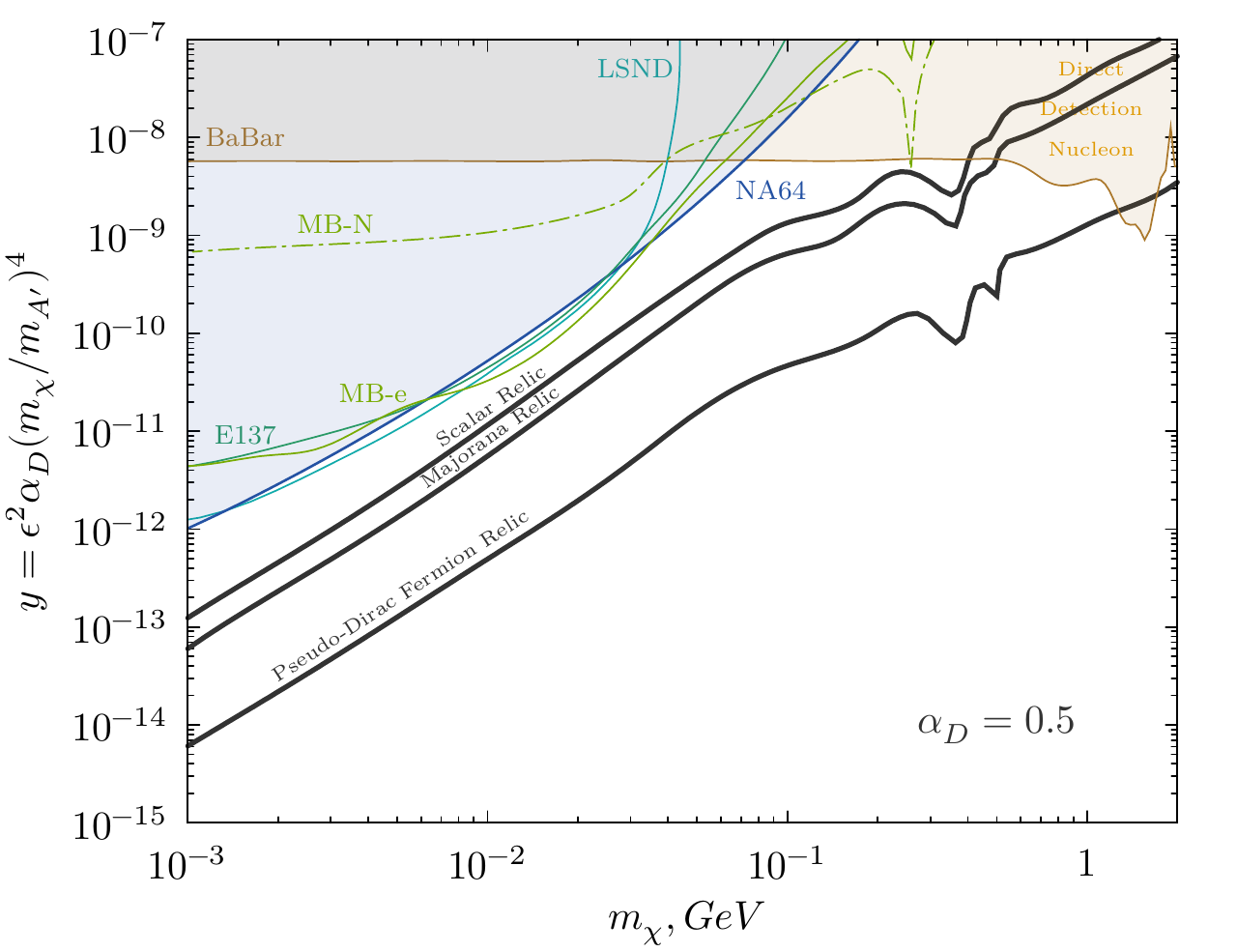}}
\includegraphics[width=.44\textwidth]{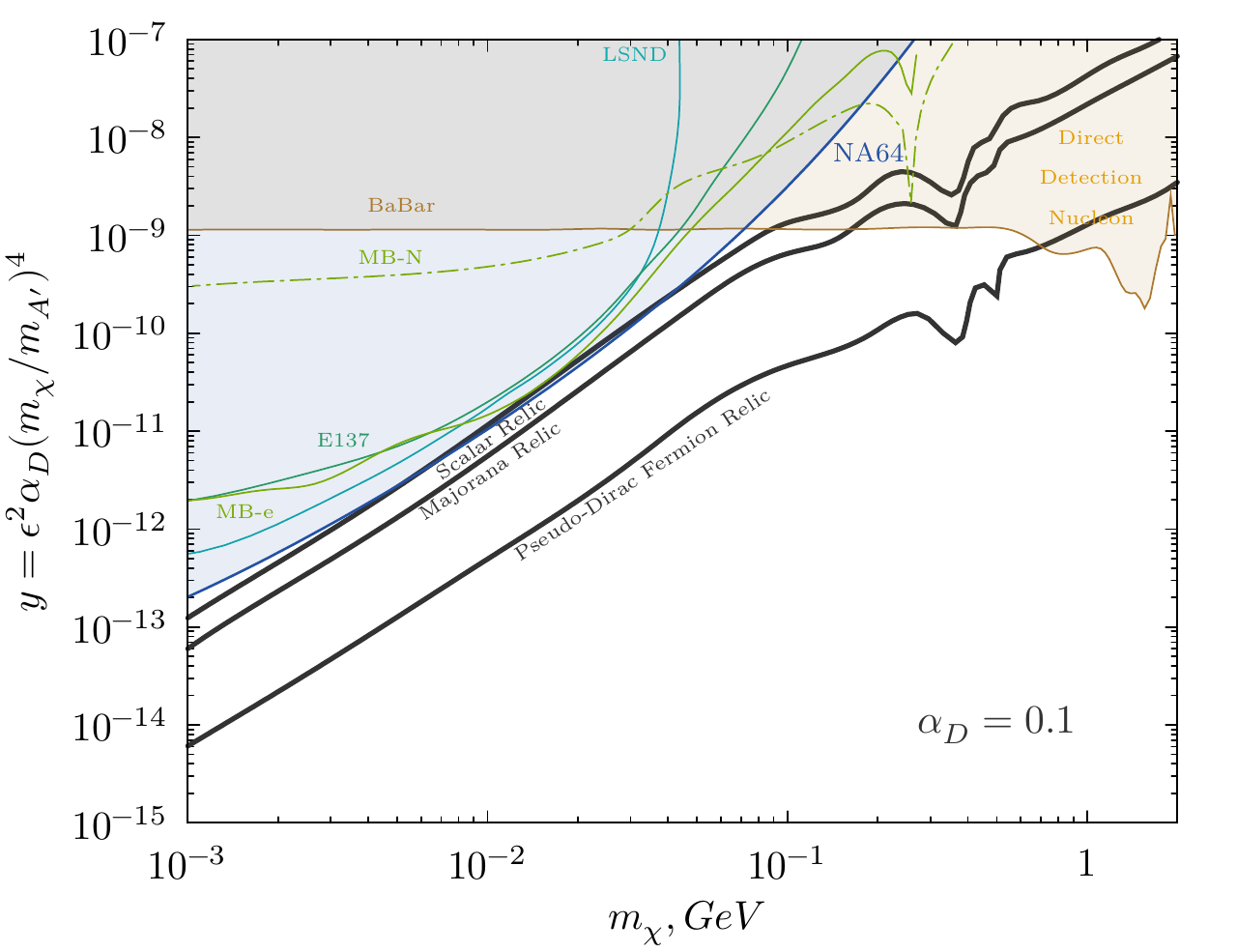}
\includegraphics[width=0.47\textwidth]{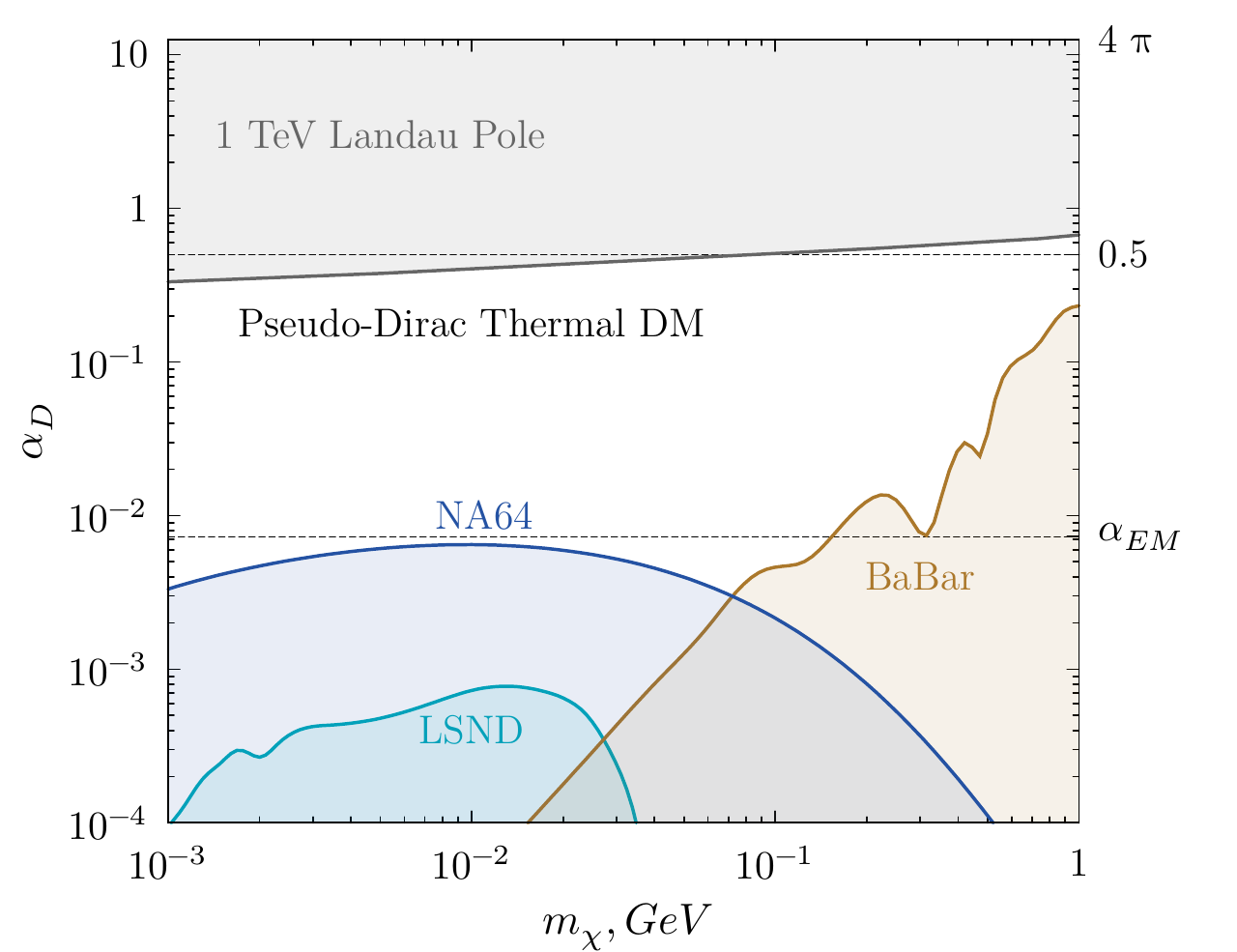}
\includegraphics[width=0.47\textwidth]{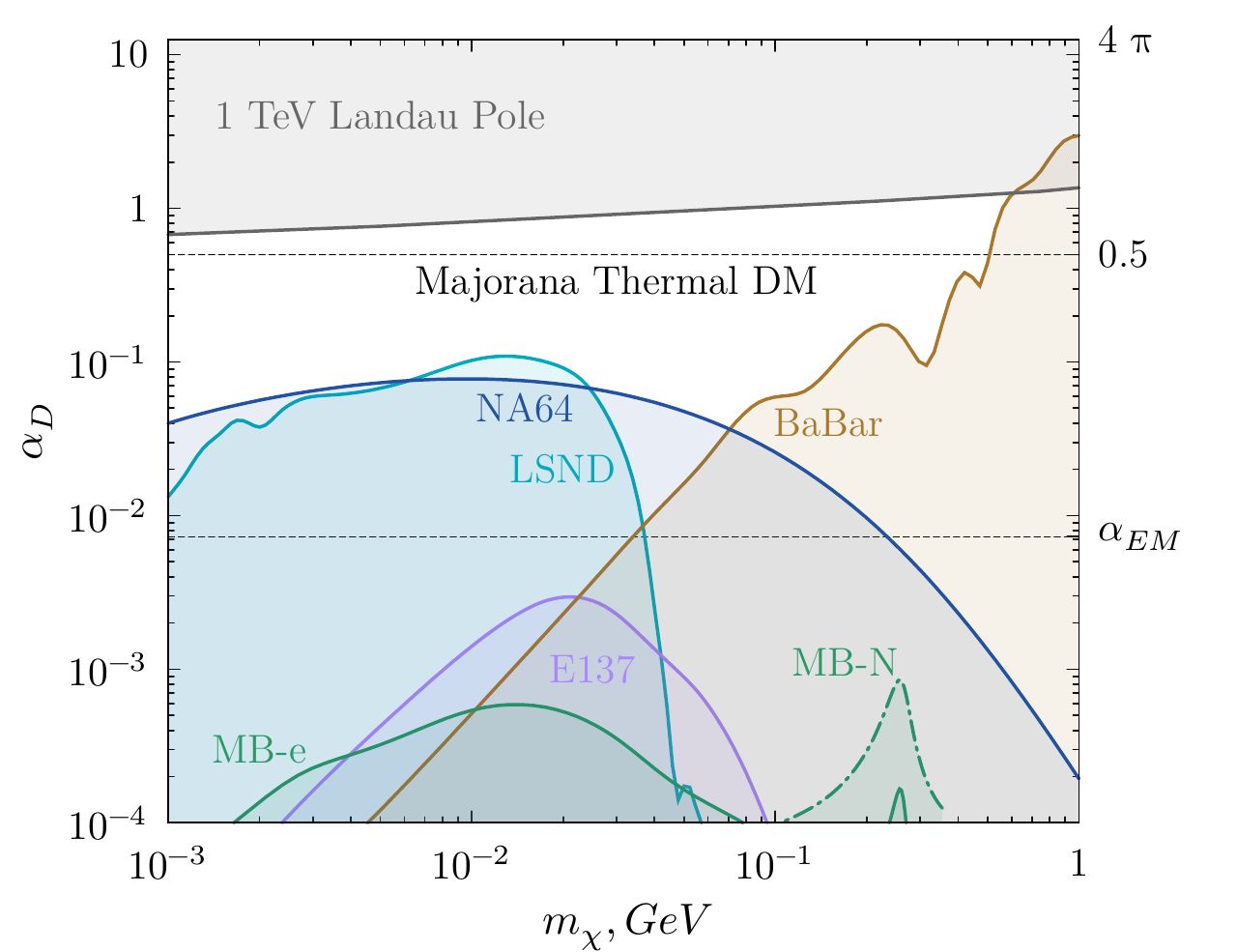}
\caption{The top row shows the NA64 limits  in  the (y;$m_{\chi}$) plane obtained for $\alpha_D=0.5$ (left panel) and $\alpha_D=0.1$ (right panel) from the full 2016-2018  data set.
The bottom row shows the NA64 constraints in the ($\alpha_D$;$m_{\chi}$) plane  on  the pseudo-Dirac (left panel) 
and Majorana (right panel) DM. The limits are shown in  
 in comparison with bounds obtained in Refs.\cite{report1, report2, Izaguirre:2014bca,Iza2015,Iza2017}  from  the results of the 
LSND~\cite{deNiverville:2011it,Batell:2009di},   E137 \cite{e137th}, MiniBooNE \cite{minib2018}, $BABAR$ \cite{babarg-2},  and direct detection \cite{mardon} experiments.
The favored parameters to account for the observed relic DM density for the  scalar, pseudo-Dirac  and Majorana  type of light DM are shown as the lowest solid line
in top plots; see, e.g. \cite{berlin}.}
\label{yvsm}
\end{figure*}   
\par  The combined 90\%  confidence level (C.L.)  upper limits for  $\epsilon$  were determined 
 by using the modified frequentist approach for confidence levels, taking the profile
likelihood as a test statistic in the asymptotic approximation \cite{junk,limit,Read:2002hq}. The total number of expected signal events in the
signal box was the sum of expected events from the three runs:
\begin{equation}
\Na = \sum_{i=1}^{3} N_{A'}^i = \sum_{i=1}^{3} n_{EOT}^i  \epsilon_{A'}^i n_{A'}^i(\epsilon,\ma, \Delta E_{e})
\label{nev}
\end{equation}
where $\epsilon_{A'}^i$ is the signal efficiency in run $i$, and $n_{A'}^i(\epsilon,\ma, \Delta E_{A'})$ is the 
signal yield per EOT generated in the energy  range $\Delta E_{e} $. 
Each $i$th entry in this sum was calculated with simulations of signal events and  processing them through 
the reconstruction program  with the same selection criteria and  efficiency corrections as for the  data sample from run $i$.
The combined 90\% C.L. exclusion limits on the mixing
strength as a function of the $A'$ mass,  calculated by taken into account the expected backgrounds and estimated systematic errors, can be seen  in Fig.~\ref{exclinv}.
The derived bounds are currently  the best for the mass range
$0.001\lesssim \ma \lesssim 0.2 $ GeV obtained from direct searches of  $\ainv$ decays \cite{pdg}.
\par  The overall signal efficiency   $\epsilon_{A'}$ is slightly $m_{A'},E_{A'}$ dependent and  is given by the 
product of efficiencies accounting for the  geometrical acceptance (0.97), the  track   ($\simeq 0.83$), SRD ($\gtrsim 0.95$), VETO ( 0.94)  and HCAL (0.94) signal reconstruction,  and  the DAQ dead time (0.93).  The signal acceptance loss due to pileup was $\simeq 8\%$ for high-intensity  runs.
The VETO and HCAL efficiency was  defined as a fraction of events below the corresponding zero-energy thresholds. The spectrum  of the energy distributions in these detectors  from the leak of the signal shower energy in the ECAL  was simulated  for different $A'$ masses \cite{gkkk} and cross-checked with measurements at the 
$e^- $ beam. The uncertainty in the VETO and HCAL efficiency for the signal events, dominated mostly by the pileup effect  from penetrating hadrons in the high-intensity  run III,  was estimated to be $\lesssim 4\%$.  The trigger efficiency  was found to be $0.95$  with a small uncertainty  2\%. The $A'$ acceptance  was evaluated by taking into account the selection efficiency for the    $e-m$ shower shape   in the ECAL from signal events \cite{gkkk}.  
The $A'$ production cross section in the primary reaction  was obtained with the exact tree-level calculations as  described in Ref.\cite{gkkketl}. 
An additional uncertainty in the $A'$ yield $\simeq 10\%$ was conservatively  accounted for  the difference between the predicted and measured dimuon yield \cite{na64prl, na64prd}, which was  the dominant source of  systematic uncertainties  on the expected number of signal events. 
The total signal  efficiency $\epsilon_{A'}$  for high- (low-) intensity runs varied from   0.53$\pm$ 0.09 (0.69$\pm$0.09) 
 to 0.48$\pm$0.08 (0.55$\pm$0.07) decreasing for the  higher  $A'$ masses. 
\par Using  constraints on the cross section of  the DM annihilation freeze-out
 [see Eq.(\ref{alphad})], and obtained limits on mixing 
strength, one  can derive constraints on the LDM models, which are shown 
 in the ($y$;$m_{\chi}$) and ($\alpha_D$;$m_{\chi}$)  planes in Fig.~\ref{yvsm} for masses  $m_{\chi} \lesssim 1$~GeV.  On the same plot one can also see   
the favored $y$ parameter curves  for scalar,  pseudo-Dirac (with a small splitting)  and Majorana  scenario  of  LDM  obtained by  taking into account  the observed relic DM density \cite{berlin}. The limits on the variable $y$  are calculated 
   under the convention $\alpha_D= 0.1$ and 0.5, and  $m_{A'}=3 m_{\chi}$ \cite{report2, pbc-bsm} and shown also for comparison with bounds  from other experiments. This choice of the $\alpha_D$ region  is compatible with the bounds 
   derived  based on the running of the dark gauge coupling arguments of Refs. \cite{gkkketl,davou}.   
     It  should be 
 noted  that for  smaller values of $\alpha_D$ the  NA64  limits  will be stronger, due to the fact that 
   the  signal rate in our case scales as $\epsilon^2$, instead of  $\epsilon^4 \alpha_D$ as for beam dump searches.
 The  bounds on $\alpha_D$ for  the case of pseudo-Dirac fermions  shown in Fig.~\ref{yvsm} (left panel in the bottom row)  were calculated 
  by taking  the value $f=0.25 $,  while for the Majorana case (right panel) the value   $f=3$  in Eq.(\ref{alphad}) \cite{na64prd} was used \footnote{We have made the calculations based on semianalytical
formulae of Ref.\cite{Kolb} and found that for pseudo-Dirac (Majorana)  fermions $f = 0.3 -0.4 ( 4.2 -5.3)$
for the mass range $1 \leq m_{A^`} \leq 100$  MeV. For limit calculations shown in  Fig. \ref{yvsm}  we used the conservative estimate with 
$f = 0.25 (3)$ similar to Refs. \cite{report2, na64prd}.}.
  One can see that using the NA64 approach allows us  to  obtain  more stringent bounds on $\epsilon,~y, ~ \alpha_D$ 
   for the mass range $m_{\chi} \lesssim 0.1$ GeV than the limits obtained from the results of  classical beam dump experiments, thus,  demonstrating its power for the dark matter search. 
 Further improving of the sensitivity  is  expected after  the   NA64 detector upgrade.\\
 We gratefully acknowledge the support of the CERN management and staff 
and the technical staffs of the participating institutions for their vital contributions. 
 This work was supported by the  Helmholtz-Institut f\"ur Strahlen-und Kernphysik (HISKP), University of Bonn (Germany), Joint Institute for Nuclear Research (JINR) (Dubna), the Ministry of Science and Higher Education (MSHE)  and RAS (Russia), ETH Zurich and SNSF Grant No. 169133 (Switzerland), and FONDECYT Grants  No.1191103, No. 190845, and No. 3170852, UTFSM PI~M~18~13,  and Basal Grant No. FB0821 CONICYT (Chile).

\end{document}